\documentclass[twocolumn,prd,
 reprint,
nofootinbib,
 amsmath,amssymb,
 aps,
]{revtex4}

\usepackage{comment}

\usepackage{color}
\usepackage{framed, color}
\definecolor{shadecolor}{rgb}{0.85,0.9,0.9} % definiert mir die Schattenboxen
\usepackage{dsfont}
\usepackage{graphicx}% Include figure files
\usepackage{dcolumn}% Align table columns on decimal point
\usepackage{bm}% bold math
\usepackage{rotating}
\begin{document}

%\rotatebox[origin=c]{180}{$\Gamma$}    for the 180 degree rotation of a symbol

%\includecomment{commenta}
\excludecomment{commenta}

\preprint{APS/123-QED}

\title{Diagnostics for insufficiencies of posterior calculations\\in Bayesian signal inference}%An error-diagnostic validation method for posterior distributions\\ in Bayesian signal inference}

\author{Sebastian Dorn\textsuperscript{1,2,}\footnote{sdorn@mpa-garching.mpg.de}, Niels Oppermann\textsuperscript{1}, Torsten A. En$\ss$lin\textsuperscript{1}}
\affiliation{\textsuperscript{1} Max-Planck-Institut f\"ur Astrophysik, Karl-Schwarzschild-Str.~1, D-85748 Garching, Germany\\
\textsuperscript{2} Technische Universit\"at M\"unchen, Arcisstra\ss e 21, D-80333 M\"unchen, Germany }
\date{\today}

\begin{abstract}
{\textbf{Abstract.}} We present an error-diagnostic validation method for posterior distributions in Bayesian signal inference, an advancement of a previous work. It transfers deviations from the correct posterior into characteristic deviations from a uniform distribution of a quantity constructed for this purpose. We show that this method is able to reveal and discriminate several kinds of numerical and approximation errors, as well as their impact on the posterior distribution. For this we present four typical analytical examples of posteriors with incorrect variance, skewness, position of the maximum, or normalization. We show further how this test can be applied to multidimensional signals.

\bigskip
\noindent DOI: 10.1103/PhysRevE.88.053303 \hspace{3.8cm}PACS number(s): 05.10.--a, 02.70.--c 
%\noindent\textbf{Subject headings:} critical test - Bayesian inference - software validation - error diagnostics.
\end{abstract}

\maketitle

%\tableofcontents

%____________________________________________________________________________________________________________________________
%\section{\label{sec:level1}Introduction}
\section{Introduction} 
Bayesian inference methods are gaining importance in many areas of physics, like, e.g.,~precision cosmology \cite{2012arXiv1212.5225B, 2013arXiv1303.5076P}. Dealing with Bayesian models means to grapple with the posterior probability distribution, whose calculation and simulation is often highly complex and therefore prone to errors. Rather than taking the correctness of the numerical implementation of the posterior for granted, one should validate it in some way. 

\bigskip Although there are validation approaches (e.g.,~\cite{doi:10.1198/016214504000001132,Cook06validationof}), these provide little diagnostics for the type of error. However, this information would be very useful in order to locate a mistake in a posterior calculating code or in its mathematical derivation. Therefore we introduce an advancement of a validation method developed by Cook \textit{et al.}\ \cite{Cook06validationof} that is able to detect errors in the numerical implementation as well as in the mathematical derivation. We show that the typical deviation of a quantity constructed for this purpose from a uniform distribution encodes information on the kind and intensity of errors made.

%\bigskip The remainder of this paper is organized as follows. In Sec.~II we introduce the validation method within the framework of Bayesian statistics. In Sec.~III we show the ability of the %validation approach to determine the kind of error by applying it to typical error examples and demonstrate this effect numerically in Sec.~IV. We summarize our results in Sec.~V.

%____________________________________________________________________________________________________________________________
%____________________________________________________________________________________________________________________________

\section{Posterior validation in one dimension}
\subsection{Validation approach}
Within this work we assume a data set $d$ is given in the form $d=(d_1,d_2,\dots,d_m)^T\in \mathds{R}^m$, where $m \in \mathds{N}$, and we want to extract a physical quantity, $s\in\mathds{R}$, from its posterior probability density function (PDF), $P(s|d)$. The data are drawn from the likelihood $P(d|s)$,

\begin{equation}
\label{basic}
d\hookleftarrow P(d|s).
\end{equation}

\noindent The posterior is given by Bayes' Theorem \cite{Bayes01011763},

\begin{equation}
\label{defZ}
P(s|d)=\frac{P(d,s)}{P(d)}= \frac{P(d|s)P(s)}{P(d)},
\end{equation}
 
\noindent where the prior is denoted by $P(s)$ and the evidence by $P(d)$. A concrete example of such a calculation including approximations that require validation can be found in \cite{paper1}.

\bigskip Now we introduce the foundation of the \textit{Diagnostics for Insufficiencies of Posterior calculations} (DIP). This is a validation method for the numerical calculation of the posterior $P(s|d)$, first developed by Cook \textit{et al.}\ \cite{Cook06validationof}.
For this purpose we use the following procedure:

\begin{enumerate}
	\item Sample $s_{\text{gen}}$ from the prior $P(s)$. %from a suitable\footnote{suitable with respect to a possible approximation in the derivation of the posterior} interval $I=\left[s_a,s_b\right]$ with respect to
	\item Generate data $d$ for $s_{\text{gen}}$ according to $P(d|s_\text{gen})$.
	\item Calculate a posterior curve for given data by determining $\tilde{P}(s|d)$ according to Eq.~(\ref{defZ}), where $\tilde{P}$ denotes the posterior including possible approximations.
	\item Calculate the posterior probability for $s\leq s_\text{gen}$ according to 
	
\begin{equation}
x:=\int_{-\infty}^{s_{\text{gen}}} ds~ \tilde{P}(s|d) ~\in \left[0,1\right]
\end{equation}  
by the use of a numerical integration technique. 
	\item If the calculation of the posterior was correct, the distribution for $x$, $P(x)$, should be uniform between 0 and 1.
\end{enumerate}
The uniformity of $P(x)$ can then be checked numerically by going through steps 1--4 repeatedly. Note that the distribution of $x$ can be uniform even if there is an error in the implementation or mathematical derivation. The reason for this is the unlikely possibility of at least two errors compensating each other exactly. However, this is a fundamental problem of nearly every numerical validation method.

\medskip We show in Appendix A analytically that $P(x)=1$ if $\tilde{P}(s|d)=P(s|d)$, as an alternative to the discussion in \cite{Cook06validationof}.

%____________________________________________________________________________________________________________________________
\subsection{Diagnostics for insufficiencies of posterior calculations (DIP) in one dimension}
Here, we introduce the DIP, an error-diagnostic, graphical validation method. It is a substantial advancement of the method pointed out in Sec.\ II.A, not only able to detect errors of the posterior distribution but also their nature and their impact on calculations using the tested posterior. The DIP test is demonstrated with four typical scenarios below. Although we use Gaussians in these examples, we would like to point out that similar effects can be expected for non-Gaussian PDFs. In fact, any one-dimensional posterior can be mapped to a Gaussian distribution by a suitably constructed transformation \cite{casella2002statistical} as shown in Appendix B.

\subsubsection{Typical analytic scenarios of insufficient posteriors}
To investigate the influence of an insufficient posterior on the distribution $P(x)$ we study as an example a Gaussian posterior,
\begin{equation}
P(s|d)=\mathcal{G}{\left(s_d,\sigma^2\right)}:=\frac{1}{\sqrt{2\pi \sigma^2}}\exp{\left(-\frac{s_d^2}{2\sigma^2}\right)},
\end{equation}
\noindent with $s_d=s-\bar{s}_d$ and $\bar{s}_d$ the data-dependent maximum of the posterior.
\begin{commenta}
(see Appendix A for details).
\end{commenta}
In the following we assume the variance to be data independent and consider a wrongly determined value $x^\epsilon=\int_{-\infty}^{s_{\text{gen}}} ds~ P^\epsilon(s|d)$, where $P^\epsilon(s|d)$ is Gaussian with wrong variance or nonzero skewness or wrong maximum position or wrong normalization.

\begin{figure}[t]
\includegraphics[width=\columnwidth]{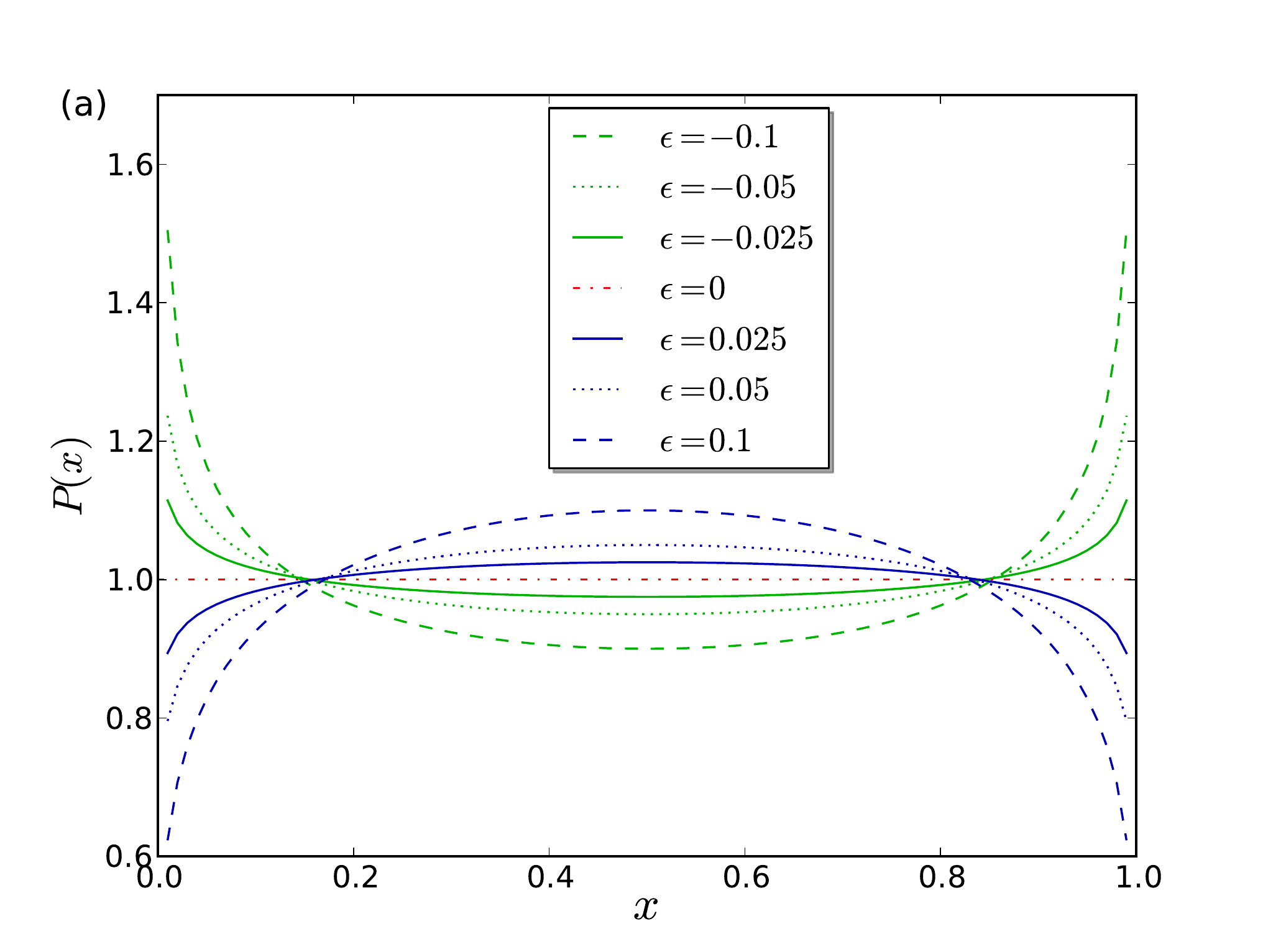}
\includegraphics[width=\columnwidth]{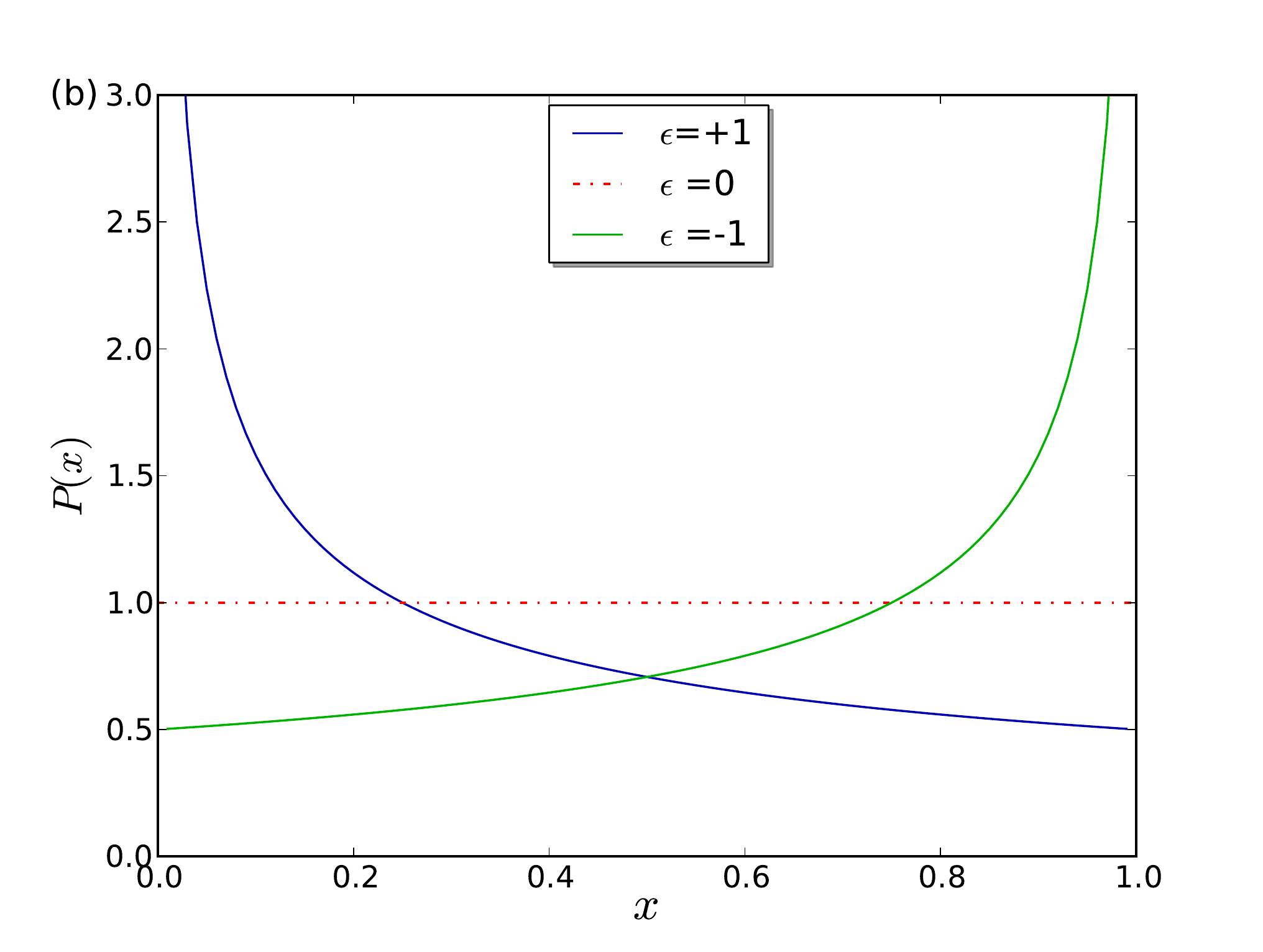}
\caption[width=\columnwidth]{(Color online) Influence of an insufficient posterior on the DIP distribution $P(x)$. The (a) upper [(b) lower] panel shows the effect of calculating $P(x)$ from a posterior with wrong variance [skewness] as described by Eq.~(\ref{wrongvar2}) [Eq.~(\ref{wrongvar3})].}
\label{wrong}
\end{figure}

\bigskip\textit{Wrong variance.} In the case in which $P(x)$ was calculated from a posterior whose standard deviation deviates by a fraction $\epsilon$ from the true value of $\sigma$, we consider

\begin{equation}
\label{wrongvar2}
\begin{split}
P^\epsilon(s|d)&=%\mathcal{G}^\epsilon(s_d,\sigma^2)=
\frac{1}{\sqrt{2\pi}\sigma(1+\epsilon)}\exp{\left(-\frac{s^2_d}{2\sigma^2(1+\epsilon)^2}\right)},\\
x^\epsilon &= \frac{1}{2}\left[1+\text{erf}{\left(\frac{s_d}{\sqrt{2}\sigma(1+\epsilon)}\right)}\right]
\end{split}
\end{equation} 

\noindent with $\epsilon>-1$. To determine the distribution $P(x)$ we use Eq.~(\ref{proof}). This yields
\begin{commenta}
(see Appendix A)
\end{commenta}

\begin{equation}
\label{wvar}
P(x)=(1+\epsilon)\exp{\left(-\left[\text{erf}^{-1}{\left(2x -1 \right)}\right]^2 \left[(1+\epsilon)^2 - 1\right]\right)}
\end{equation}

\noindent with the limit $P(x)\stackrel{\epsilon\rightarrow 0}{\longrightarrow}1$. The deviations from the uniform distribution increase with the value of $|\epsilon|$ and are shown in Fig.~\ref{wrong}(a). In case the standard deviation was underestimated, $\epsilon<0$, the distribution for $x$ becomes convex (``$\cup$ shape") and for an overestimation, $\epsilon>0$, it becomes concave (``$\cap$ shape"). Since the underestimation of variances is a typical mistake, the DIP test produces often test distributions with a dip in the middle. %This means that the standard deviation interval around the expectation value (maximum of the Gaussian PDF) is too small (convex) or too big (concave).

\bigskip\textit{Wrong skewness.} Next, we consider the case in which $P(x)$ was calculated from a falsely skewed posterior,
\begin{equation}
\label{wrongvar3}
P^\epsilon(s|d)=%\mathcal{G}^\epsilon(s_d,\sigma^2)=
\frac{1}{\sqrt{2\pi}\sigma}\exp{\left(-\frac{s^2_d}{2\sigma^2}\right)}\left(1+ \text{erf}{\left(\frac{\epsilon s_d}{\sqrt{2}\sigma}\right)}\right).
\end{equation} 

\noindent Thus, $x^\epsilon$ is given by

\begin{commenta}
\begin{equation}
\label{abc}
\begin{split}
x^\epsilon=&\frac{1}{2}\left[1+\text{erf}{\left(\frac{s_d}{\sqrt{2}\sigma}\right)}\right]\\
	 &+\frac{1}{\sqrt{2\pi}\sigma}\int_{-\infty}^{s_d} d\tilde{s}_d ~\exp{\left(-\frac{\tilde{s}_d^2}{2\sigma^2}\right)}~\text{erf}\left(\frac{\epsilon \tilde{s}_d}{\sqrt{2}\sigma}\right).
\end{split}
\end{equation}

\noindent Now we insert the identity

\begin{equation}
\begin{split}
\text{erf}\left(\frac{\epsilon \tilde{s}_d}{\sqrt{2}\sigma}\right)=&\int_0^\epsilon d\tilde{\epsilon}~\frac{\partial}{\partial \tilde{\epsilon}}~\text{erf}\left(\frac{\tilde{\epsilon} \tilde{s}_d}{\sqrt{2}\sigma}\right)\\
 =& \sqrt{\frac{2}{\pi}} \int_0^\epsilon d\tilde{\epsilon}~\frac{\tilde{s}_d}{\sigma}~\exp\left(-\frac{\tilde{\epsilon}^2 \tilde{s}_d^2}{2\sigma^2}\right)
\end{split}
\end{equation}

\noindent into Eq.~(\ref{abc}) and perform the $\tilde{s}_d$-integration first to obtain
\end{commenta}

\begin{equation}
\label{13}
\begin{split}
x^\epsilon=&\frac{1}{2}\left[1+\text{erf}{\left(\frac{s_d}{\sqrt{2}\sigma}\right)}\right]\\
	&-\frac{1}{\pi}\int_0^\epsilon d\tilde{\epsilon}~\frac{\exp{\left(-\frac{1}{2}\left(\frac{s_d}{\sigma}\right)^2 \left(1+\tilde{\epsilon}^2\right) \right)}}{1+\tilde{\epsilon}^2}\\
	=:&\frac{1}{2}\left[1+\text{erf}{\left(\frac{s_d}{\sqrt{2}\sigma}\right)}\right] -2T{\left(\frac{s_d}{\sigma},\epsilon\right)},
\end{split}
\end{equation}

\noindent where $T{\left(\frac{s_d}{\sigma},\epsilon\right)}$ is the \textit{Owen's function} \cite{Owen}, and $\epsilon$ denotes the dimensionless skewness parameter. Now we focus on $|\epsilon|=1$ for simplicity, for which $2T{\left(\frac{s_d}{\sigma},\pm 1\right)} = \pm \frac{1}{4}\left(1-\text{erf}^2{\left(\frac{s_d}{\sqrt{2}\sigma}\right)}\right)$. Applying Eq.~(\ref{proof}) yields
\begin{commenta}
(see Appendix A)
\end{commenta}

\begin{equation}
\label{wskew}
P(x)=\left\{
    		\begin{array}{cc}
                		 \left(2\sqrt{x}\right)^{-1} &~~~~~~~~~~\text{if}~ \epsilon=1\\
                 		 \left(2\sqrt{1-x}\right)^{-1}&~~~~~~~~~~\text{if}~ \epsilon=-1
    		\end{array} 
    		\right..
\end{equation}

\noindent The effect of an incorrectly skewed posterior is an enhancement of values close to $x=0$ or $x=1$ [Fig.~\ref{wrong}(b)] and means that the $68\%$ confidence interval around the expectation value (maximum of the Gaussian PDF) is falsely calculated to be asymmetric. Here, we restricted ourselves to the cases $\epsilon = \pm 1$ due to their analytic treatability. Smaller deviations with $|\epsilon|<1$ will lead to qualitatively similar but less pronounced distortions of the sampled distribution $P(x)$.

\bigskip\textit{Wrong maximum position.} In the case in which $P(x)$ was calculated from a posterior whose maximum has a wrong position, we consider 
\begin{equation}
\label{maxpos}
\begin{split}
P^\epsilon(s|d)&=%\mathcal{G}^\epsilon(s_d,\sigma^2)=
\frac{1}{\sqrt{2\pi}\sigma}\exp{\left(-\frac{(s_d-\epsilon)^2}{2\sigma^2}\right)},\\
x^\epsilon &= \frac{1}{2}\left[1+\text{erf}{\left(\frac{s_d - \epsilon}{\sqrt{2}\sigma}\right)}\right].
\end{split}
\end{equation} 

\begin{figure}[t]
\includegraphics[width=\columnwidth]{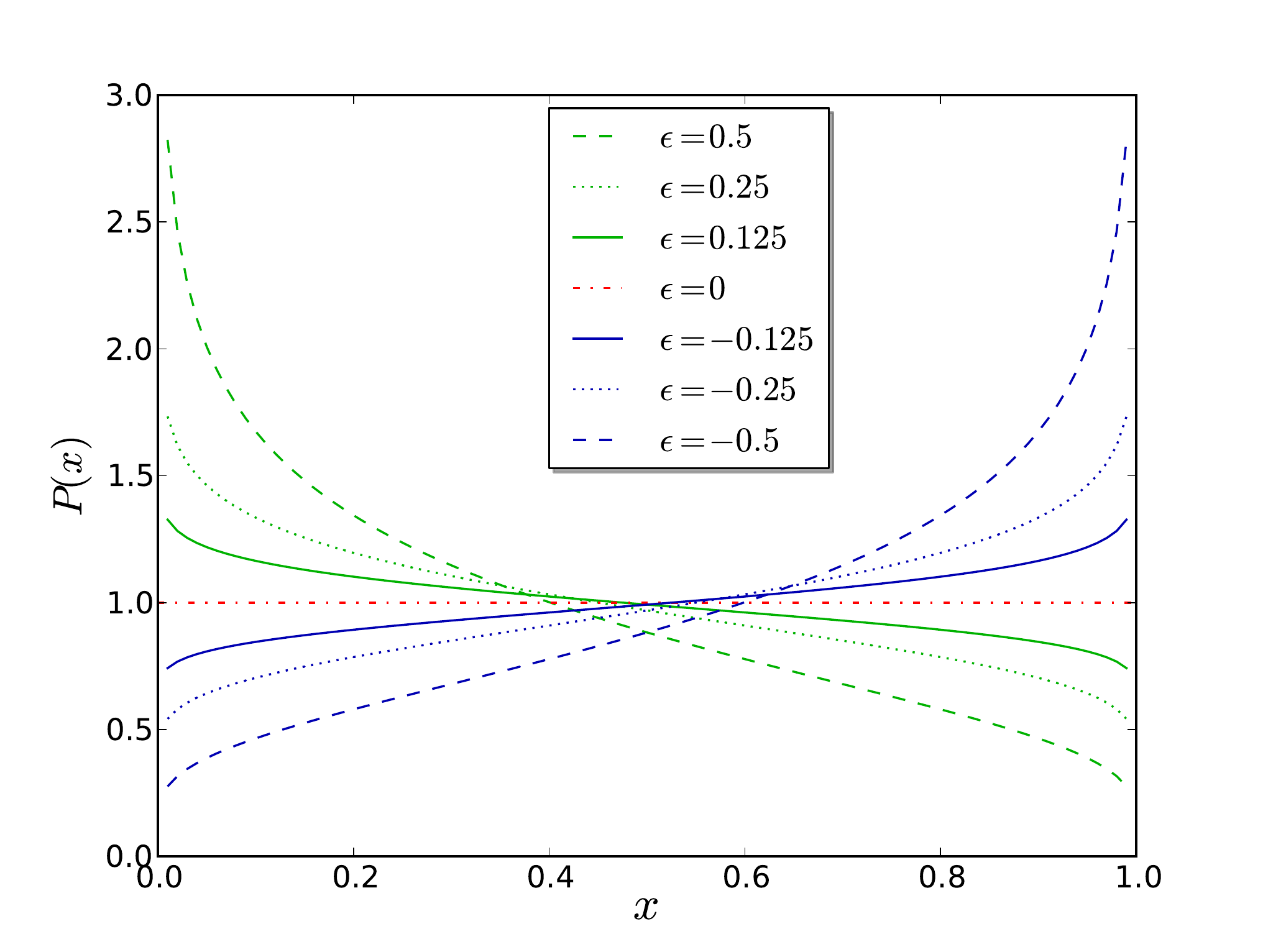}
\caption[width=\columnwidth]{(Color online) Influence of an insufficient posterior on the DIP distribution $P(x)$. The panel is showing the effect of calculating $P(x)$ from a posterior with wrong maximum position as described by Eq.~(\ref{maxpos}).}
\label{wrongmax}
\end{figure}

\noindent Applying again Eq.~(\ref{proof}) yields
\begin{commenta}
(see Appendix A)
\end{commenta}

\begin{equation}
\label{wmax}
P(x)=\exp{\left(-\frac{1}{2}\left(\frac{\epsilon}{\sigma}\right)^2 -\sqrt{2}\left(\frac{\epsilon}{\sigma}\right)~\text{erf}^{-1}{\left(2x -1\right)}\right)},
\end{equation}

\noindent with the limit $P(x)\stackrel{\epsilon\rightarrow 0}{\longrightarrow}1$. The resulting distribution of $x$ for $\sigma=1$ is shown in Fig.~\ref{wrongmax}. Here, the $x$ abundances near $x=0$ or $x=1$ are enhanced, similarly to the case of incorrect skewness. However, the slope of $P(x)$ at the suppressed end differs significantly from the former case.

\bigskip\textit{Wrong normalization.} Lastly, we consider the case in which $P(x)$ was calculated from a posterior with wrong normalization,
\begin{equation}
\label{norm}
\begin{split}
P^\epsilon(s|d)&=%\mathcal{G}^\epsilon(s_d,\sigma^2)=
\frac{1}{\sqrt{2\pi}\sigma (1+\epsilon)}\exp{\left(-\frac{s_d^2}{2\sigma^2}\right)},\\
x^\epsilon &= \frac{1}{2(1+\epsilon)}\left[1+\text{erf}{\left(\frac{s_d}{\sqrt{2}\sigma}\right)}\right],
\end{split}
\end{equation} 

\noindent which yields [Eq.~(\ref{proof})]
\begin{commenta}
(see Appendix A)
\end{commenta}

\begin{equation}
\label{wnorm}
P(x)=1+\epsilon~~~~~~\text{for}~x\in[0,1-\epsilon].
\end{equation}

\noindent This means the value of $\epsilon$ can be determined precisely from the $x$ interval.

\begin{commenta}
\noindent The resulting distribution of $x$  is shown in Fig.~\ref{wrongnorm}.
\begin{figure}[h]
\includegraphics[width=\columnwidth]{wrong_posterior_norm.pdf}
\caption[width=\columnwidth]{(Color online) Influence of an insufficient posterior on the distribution $P(x)$. The panel is showing the effect of calculating $P(x)$ from a posterior with wrong normalization.}
\label{wrongnorm}
\end{figure}
\end{commenta}

%____________________________________________________________________________________________________________________________
\subsubsection{DIP -- overview}
Table \ref{table1} summarizes typical error signatures that can easily be detected by visual inspection of the DIP test.%our findings and demonstrates the practical preferences of the DIP test. The latter are given by the fact that one can check the accuracy of a posterior distribution or detect the type of error and its impact on calculations using the posterior by eye in a very uncomplicated and fast way.%
\begin{table}[t]
\caption{\label{table1}
Summary of the DIP test scenarios. The table shows the connection between the numerical (or caused by approximations) error type and the graphical effect.}
\begin{ruledtabular}
\begin{tabular}{l|l}
\text{Graphical effect}&
\text{Error type}\\%&
%\text{Impact}\\
\colrule
Flat distribution	& --\\			%& accurate posterior\\
\hline
``$\cup$ ($\cap$)shape"  			& Variance under-(over)estimated \\% \text{error-bars too small (big)}\\
\hline
$x=0$ ($x=1$) enhanced, \\concave 	& Too neg.\ (pos.) skewed \\%& \text{error-bars too asymmetric}\\
\hline
$x=0$ ($x=1$) enhanced, \\concave and convex 	& Too large (low) max.\ postition \\%& \text{wrong location of the error-bars }\\
\hline
$x$-interval smaller \\(greater) than one 	& Too large (low) normalization \\%& \text{probabilities too small (large)}\\
\end{tabular}
\end{ruledtabular}
\end{table}

Note that any combination of the errors mentioned in Table \ref{table1} could appear, translating into a superposition of the particular graphical effects. An asymmetric ``$\cup$ shape", for instance, where abundances near $x=0$ are slightly enhanced would illustrate an underestimation of the variance combined with a too positive skewness or a too large maximum position. However, in practice often one error dominates. In that case the fitting formulas of Sec.\ II.B.1 are applicable. 

%____________________________________________________________________________________________________________________________
\subsection{Numerical example of an insufficient posterior}
Next, we demonstrate the effects of insufficient posteriors with a numerical example. For this we generate mock data according to
\begin{equation}
\label{examp}
d = s +n,
\end{equation}

\begin{figure*}[t]
\includegraphics[width=0.66\columnwidth]{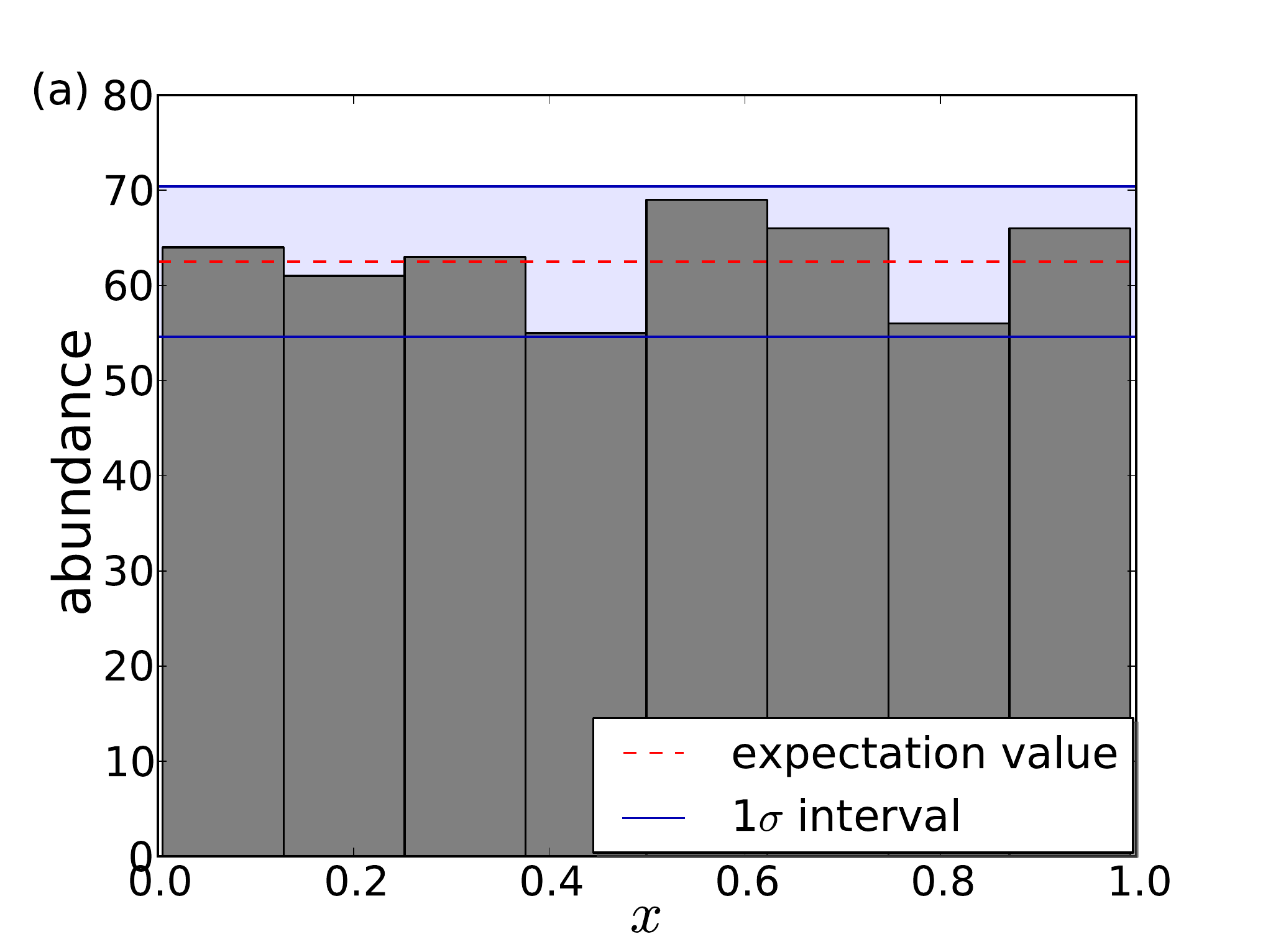}
\includegraphics[width=0.66\columnwidth]{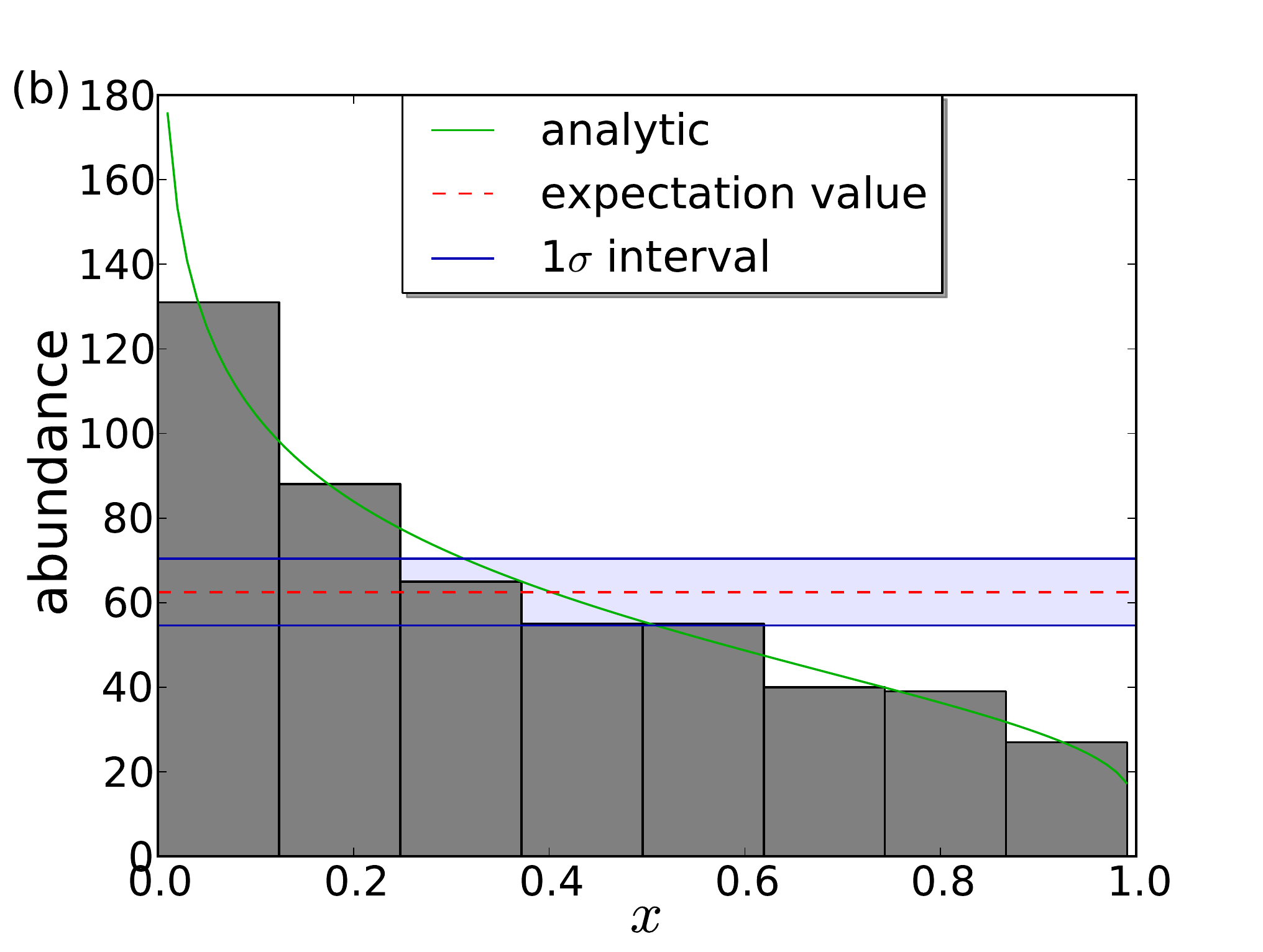}
\includegraphics[width=0.66\columnwidth]{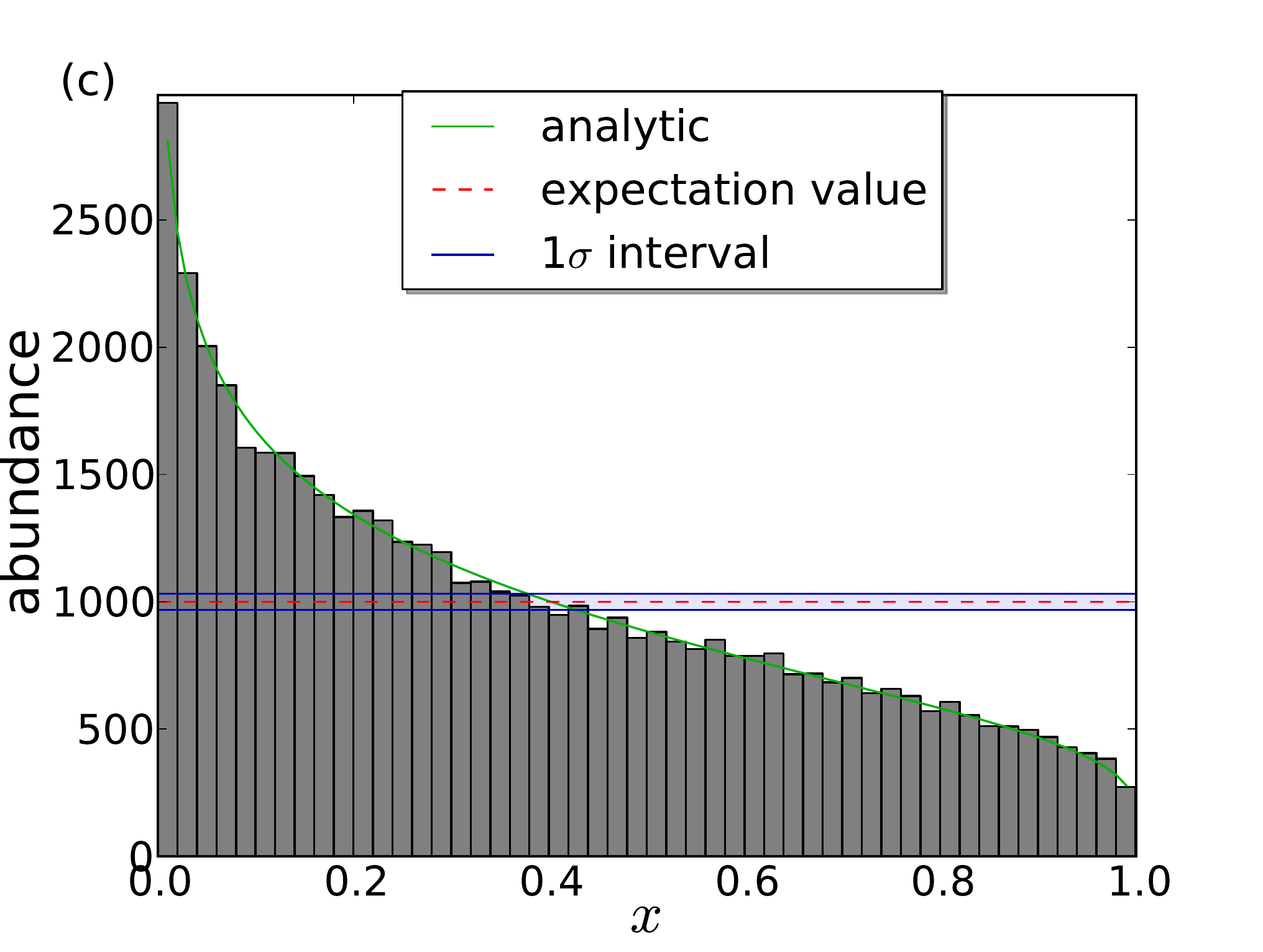}
\caption[width=\columnwidth]{(Color online) Distributions of the numerically calculated $x$-values. The (a) left [(b) middle, (c) right] histogram shows the unnormalized distribution of 500 [500, 50000] $x$-values within eight [eight, fifty] bins as calculated from the posterior with correct [wrong, wrong] maximum position. The standard deviation interval (1$\sigma$) around the expectation value as calculated from Poissonian statistics is also shown.}
\label{wrongskeweps}
\end{figure*}

\noindent where $s$ and $n$ are zero-centered Gaussian random numbers with covariance $S=1$ and $N=0.1$, respectively. To reconstruct $s$ optimally from the data we apply a \textit{Wiener filter} \cite{wiener1964time} on $d$,
\begin{equation}
m = \underbrace{\left(S^{-1} +  N^{-1} \right)^{-1}}_{=:D} N^{-1}d.
\end{equation}
\noindent After that, the posterior for $s$ is given by
\begin{equation}
\label{postex}
P(s|d) = \mathcal{G}(s-m,D).
\end{equation}

\noindent To investigate the accuracy of our implementation we go through the 
DIP validation procedure. For that purpose we sample $s_\text{gen}$ values from the distribution 
$\mathcal{G}(s,S)$. Next, we generate data according to 
Eq.~(\ref{examp}) and calculate a posterior curve according to Eq.~(\ref{postex}). 
Subsequently, we numerically determine the posterior probability for 
$s<s_\text{gen}$, which is denoted by $x$. Now this procedure is 
repeated $500$ times to sample $P(x)$.

In order to demonstrate the effect of an insufficient posterior 
we falsely include a wrong maximum position with $\epsilon=0.15$,
i.e.~our wrong test posterior is given by
\begin{equation}
P^{\epsilon=0.15}(s|d)=\mathcal{G}(s-m-0.15,D),
\end{equation}

\noindent and apply the validation procedure once again. Figure~\ref{wrongskeweps}
shows the distributions of $x$ for the correct and incorrect 
posterior.

\noindent The results are in agreement with the analytical 
considerations.

\subsection{Application to an actual physical problem}
An application of the DIP test in precision cosmology and its 
implications is given in \cite{paper1}. There, a new way to calculate the 
posterior for the local primordial non-Gaussianity parameter $f_\text{nl}$ from 
cosmic microwave background observations is presented and validated via the DIP test. Thereby a numerical problem in the implementation of the posterior could be detected and classified. 

%___________________________________________________________________________________________________________________________________________
\section{DIP in higher dimensions}
Although we have presented the DIP test in one dimension ($s\in\mathds{R}$), this approach can in principle\footnote{Note that the DIP test might become computationally expensive for high-dimensional problems.} be extended to arbitrary dimensions ($t\in\mathds{R}^m,~m\in \mathds{N}$) by mapping this multidimensional posterior $P(t|d)$ onto one dimension, $s=s(t)\in\mathds{R}$, by the usage of a marginalization, $P(s|d)=\int\mathcal{D}t~P(s|t,d)P(t,d)$. Now it is possible to apply the DIP test for the remaining coordinate, $P(s|d)$. Because there are infinitely many ways to perform the mapping, $t\mapsto s=s(t)$, a suite of tests can be constructed to probe $P(t|d)$ in various ways. A combination of these tests then yields a multidimensional posterior test. 

In the following two two-dimensional examples should illustrate this.

\subsection{Analytical example in two dimensions}
The following analytical example should demonstrate the DIP test in higher dimensions. Within this example we choose mappings onto one dimension (parametrized by $\phi$), which are not the most suitable ones to detect an error. We will show, however, that the DIP test is still able to detect and classify an error. For this purpose we assume a correct posterior distribution, given by a two-dimensional Gaussian with zero mean,

\begin{equation}
\begin{split}
P&(t|d)= \frac{1}{\sqrt{(2\pi)^2}\sigma_x\sigma_y}\\
\times & \exp\left(-\frac{1}{2}\left(\begin{array}{c} t_{x} \\ t_{y} \end{array}\right)^T \left( \begin{array}{cc} \sigma_x^{-2} & 0 \\
      0 & \sigma_y^{-2} \end{array}\right) \left(\begin{array}{c} t_{x} \\ t_{y} \end{array}\right)\right),
\end{split}
\end{equation}

\noindent and falsely manipulate the variance by setting $\sigma_x \rightarrow \sigma_x (1+\epsilon)$, i.e.,\ we consider a wrong distribution, $P^\epsilon(t|d)$, with too large standard deviation along the $t_{x}$ axis. From now on we set $\sigma_x=\sigma_y=:\sigma$ for simplicity. Next, we have to map the test distribution, $P^\epsilon(t|d)$, onto one dimension to apply the DIP test. One way to do this is to consider the  intersection of $P^\epsilon(t|d)$ with the hypersurface given by $t_{y}=t_{x}\tan(\phi)$, where $\phi\in[0,2\pi]$ denotes the usual azimuth in the $t_{y}$-$t_{x}$ plane. After this mapping (and choice of a proper normalization) we obtain

\begin{equation}
\begin{split}
P_\text{1d}(t|d) = &~ \frac{1}{\sqrt{2\pi}\sigma_\phi}~\exp\left(-\frac{1}{2}\frac{t_{x}^2}{\sigma_\phi^2}\right),\\
P_\text{1d}^\epsilon(t|d) = &~ \frac{1}{\sqrt{2\pi}\sigma_\phi^\epsilon}~\exp\left(-\frac{1}{2}\frac{t_{x}^2}{\left(\sigma_\phi^\epsilon\right)^2}\right),
\end{split}
\end{equation}

with 

\begin{equation}
\begin{split}
\sigma_\phi := & ~\frac{\sigma}{\sqrt{1+\tan^2(\phi)}},\\
\sigma_\phi^\epsilon := &~ \frac{\sigma}{\sqrt{\frac{1}{(1+\epsilon)^2}+\tan^2(\phi)}}.
\end{split}
\end{equation}

The determination of the $\phi$ dependent $P(x)$, $P_\phi(x)$, works analogous to the \textit{wrong variance} section of II.B.1 and yields

\begin{equation}
\label{2deq}
P_\phi(x) = \frac{\sigma_\phi^\epsilon}{\sigma_\phi}\exp\left(-\left[\text{erf}^{-1}{\left(2x -1 \right)}\right]^2 \left[\left(\frac{\sigma_\phi^\epsilon}{\sigma_\phi}\right)^2 - 1\right]\right),
\end{equation}

\noindent with the limits

\begin{equation}
\begin{split}
\frac{\sigma_{\phi}^\epsilon}{\sigma_{\phi}}\bigg\vert_{\phi=0,\pi}= &~ 1 + \epsilon,\\
\lim_{\phi \rightarrow \frac{\pi}{2}, \frac{3\pi}{2}} \left(\frac{\sigma_\phi^\epsilon}{\sigma_\phi}\right) = &~ 1.
\end{split} 
\end{equation}

\begin{figure}[t]
\includegraphics[width=\columnwidth]{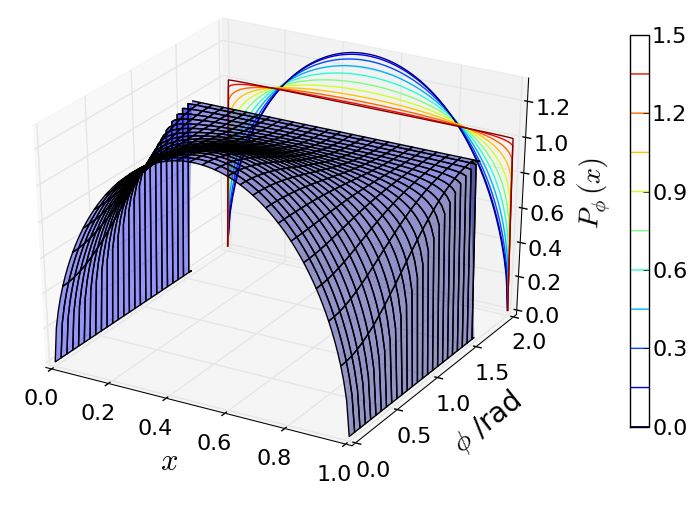}
\caption[width=\columnwidth]{(Color online) DIP test in two dimensions with related contour plot, according to Eq.\ (\ref{2deq}). $\phi$ denotes the azimuth (of the $t_{x}$-$t_{y}$ plane), where the two-dimensional posterior $P^\epsilon(s|d)$ intersects with a plane, given by $t_{y}=t_{x} \tan(\phi)$. The DIP distribution of the resulting one-dimensional posterior is denoted by $P_\phi(x)$. The labeling of the color bar refers to the $\phi$ coordinate.}
\label{2d}
\end{figure}

Figure \ref{2d} illustrates this result for an overestimation of the variance, $\epsilon=0.3$, and $\phi\in[0,\frac{\pi}{2}]$. In this case the DIP test shows significant deviations from an accurate posterior within a finite $\phi$ range. That means the DIP test indicates the insufficiency of the posterior even without hitting the exact parameter constellation (here, $\phi = 0$). Thus, in this example it is sufficient to perform at most two DIP tests (even though one can construct infinitely many) with $\Delta\phi=\frac{\pi}{2}$ to get an indication of the insufficiency of the posterior.

\subsection{Numerical example of a Bayesian hierarchical model}

\begin{figure}[t]
\includegraphics[width=.4\columnwidth]{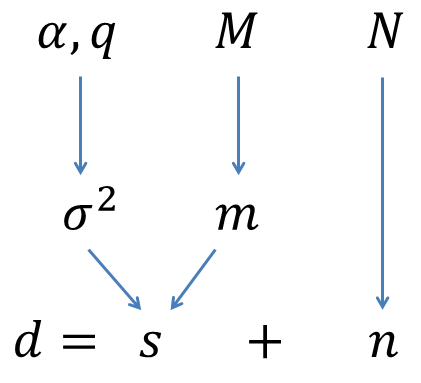}
\caption[width=\columnwidth]{(Color online) Scheme of the Bayesian hierarchical model.}
\label{scheme}
\end{figure}

To demonstrate the practical relevance in posterior computation we consider a Bayesian hierarchical model, where the data\footnote{We study a single data point for simplicity because we are just interested in the accuracy of the posterior, not its usefulness for determining $m$ and $\sigma^2$.} $d\in \mathds{R}$ are given by $d=s+n$, with $n$ a white Gaussian noise. The signal $s$ itself depends on a mean $m\in \mathds{R}$, drawn from the Gaussian $\mathcal{G}(m,M)$ with related variance $M$ and on a signal variance, $\sigma^2\in \mathds{R}$. The signal variance is drawn from an inverse-Gamma distribution,

\begin{figure*}[t]
\includegraphics[width=0.66\columnwidth]{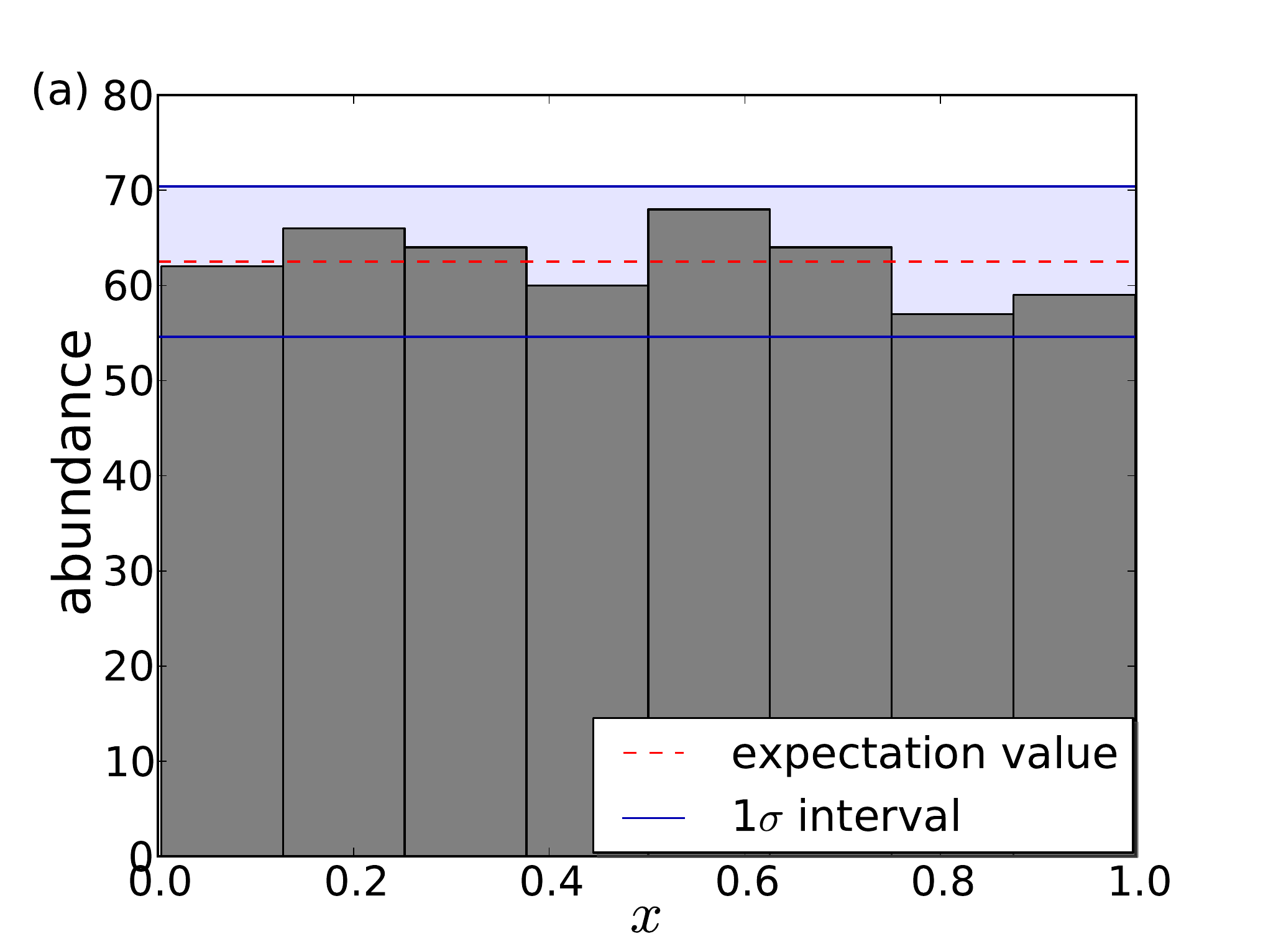}%
\includegraphics[width=0.66\columnwidth]{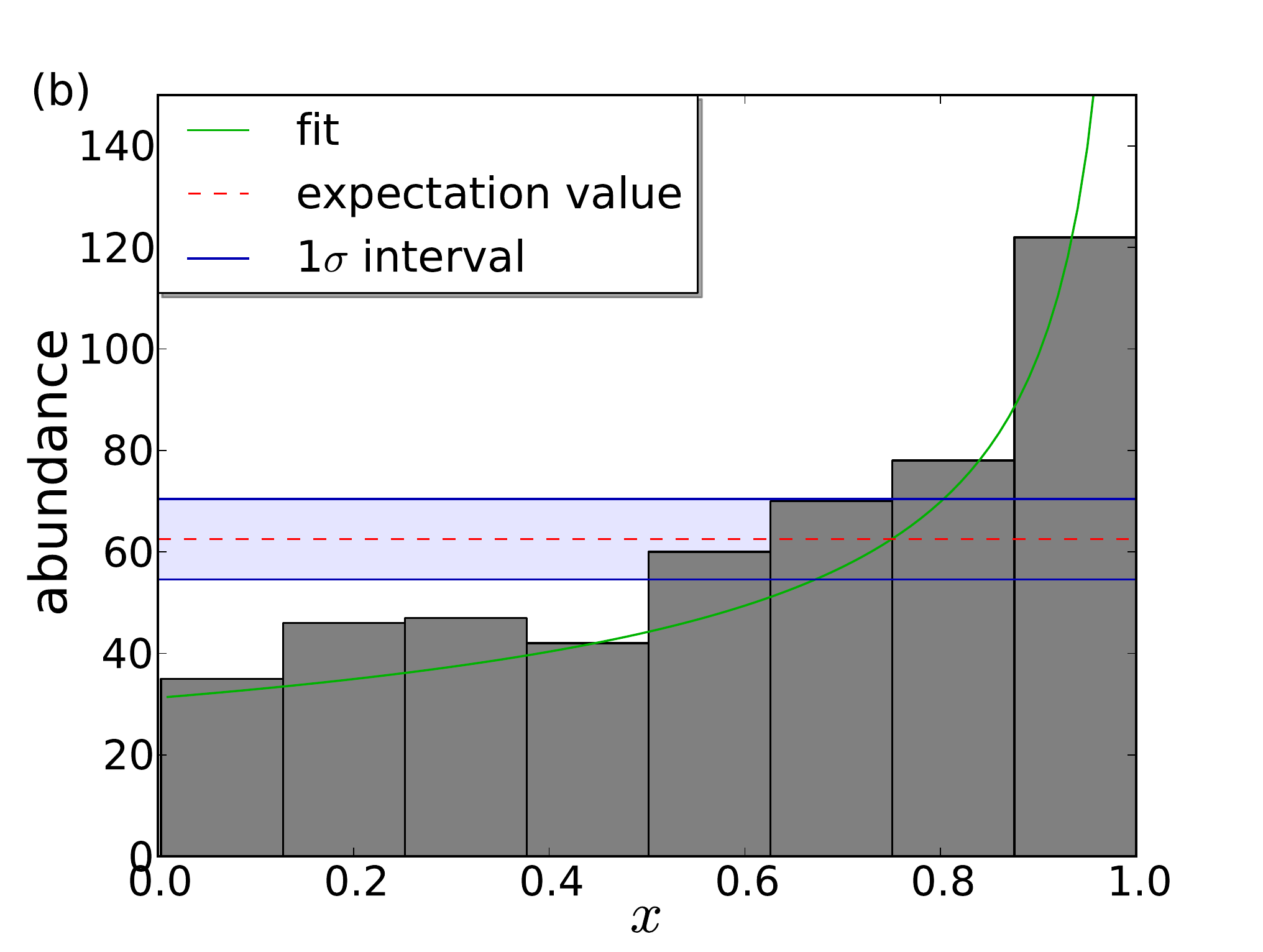}%
\includegraphics[width=0.66\columnwidth]{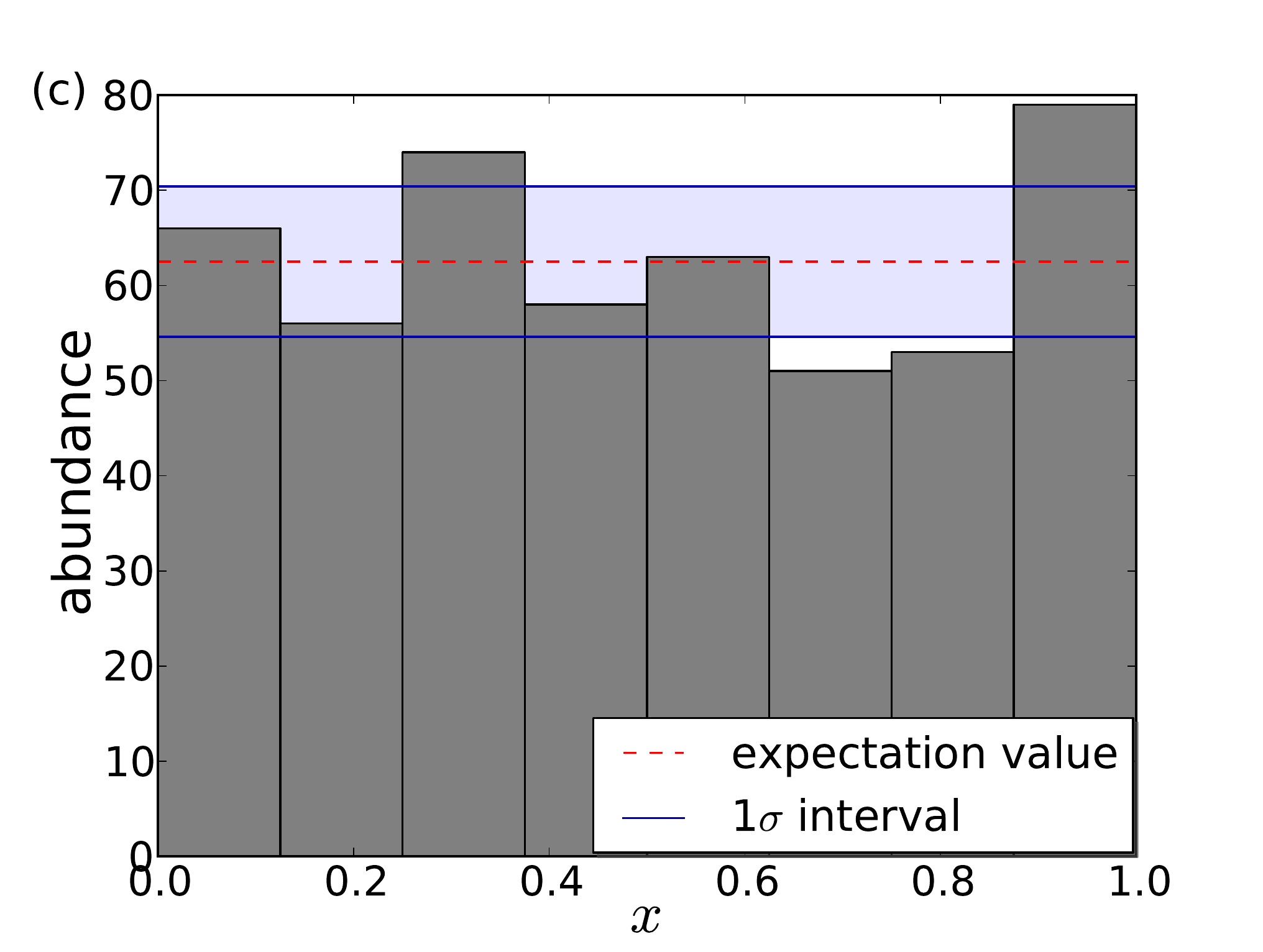}
\caption[width=\columnwidth]{(Color online) Distributions of the numerically calculated $x$ values. The (a) left [(b) middle, (c) right] histogram shows the unnormalized distribution of 500 $x$ values within eight bins as calculated from the $m$- [$m$-, $\sigma^2$-] marginalized posterior with correct [wrong, wrong] $\alpha$ parameter. The standard deviation interval (1$\sigma$) around the expectation value as calculated from Poissonian statistics is also shown.}
\label{results2d}
\end{figure*}

\begin{equation}
\mathcal{I}{\left(\sigma^2;\alpha,q\right)}:=\frac{q^\alpha}{\Gamma(\alpha)}\sigma^{2-\alpha-1}\exp{\left(-\frac{q}{\sigma^2}\right)},
\end{equation}

\noindent with $\Gamma$ the Gamma function and shape parameters $\alpha,~q$. Figure \ref{scheme} illustrates the constituents of the data.
Furthermore, we assume the following reasonable relations:
\begin{equation}
\begin{split}
P{\left(m,\sigma^2\right)} =&~ \mathcal{G}(m,M)~\mathcal{I}{\left(\sigma^2;\alpha,q\right)},\\
P{\left(s|m,\sigma^2\right)} = &~ \mathcal{G}{\left(s-m,\sigma^2\right)},~\text{and}\\
P(d|s) =&~ \mathcal{G}(d-s,N),
\end{split}
\end{equation}
\noindent where $N$ denotes the noise covariance. In this example we want to reconstruct $m$ and $\sigma^2$ from given data $d$ (a similar problem is stated in \cite{jaynes2003probability}), i.e.\ we want to determine the posterior,
\begin{equation}
\label{longpost}
\begin{split}
P&{\left(m,\sigma^2|d\right)}\\
=&~ \frac{1}{\mathcal{N}}~ \mathcal{G}(m,M)~\mathcal{I}{\left(\sigma^2;\alpha,q\right)}\\
&\times \int ds~\mathcal{G}(d-s,N)~\mathcal{G}{\left(s-m,\sigma^2\right)}\\
=&~\frac{\mathcal{G}(m,M)~\mathcal{I}{\left(\sigma^2;\alpha,q\right)}~\mathcal{G}{\left(d-m,\sigma^2 +N\right)}}{\int_0^\infty d\sigma^2 ~\mathcal{I}{\left(\sigma^2;\alpha,q\right)}~\mathcal{G}{\left(d,\sigma^2 +M+N\right)}},
\end{split}
\end{equation}
\noindent where $\mathcal{N}$ denotes the normalization. In the implementation of Eq.\ (\ref{longpost}) we falsely include an error by setting $\alpha\rightarrow \alpha(1+\epsilon)$ and apply afterwards the DIP test to investigate the accuracy of the posterior, $P^\epsilon{\left(m,\sigma^2|d\right)}$. For this purpose we have to map the posterior onto one dimension. We choose two natural mappings, given by an $m$- or $\sigma^2$ marginalization of the posterior,
\begin{equation}
\label{posta}
\begin{split}
P{\left(\sigma^2|d\right)}=&\int dm~P{\left(m,\sigma^2|d\right)}\\
 =&~\frac{1}{\mathcal{N}}~ \mathcal{G}{\left(d,\sigma^2 +M+N\right)}~\mathcal{I}{\left(\sigma^2;\alpha,q\right)},
\end{split}
\end{equation}
\noindent or
\begin{equation}
\label{postb}
\begin{split}
 P{\left(m|d\right)} =& \int_0^\infty d\sigma^2 ~P{\left(m,\sigma^2|d\right)}\\
 =&~ \frac{1}{\mathcal{N}}~ \mathcal{G}(m,M)~\\
&\times \int_0^\infty d\sigma^2 ~\mathcal{I}{\left(\sigma^2;\alpha,q\right)}~\mathcal{G}{\left(d-m,\sigma^2 +N\right)}.
\end{split}
\end{equation}
The results of the DIP test from 500 data realizations for $\alpha = 2,~q=1,~M=1,~N=0.1$ and $\epsilon=0.3$ are shown by Fig.~\ref{results2d}, where the (a) left histogram shows the accuracy of the posterior implementation for $\epsilon=0$. The (b) middle histogram shows the DIP test for $P^{\epsilon=0.3}{\left(\sigma^2|d\right)}$ according to Eq.~(\ref{posta}). Here, the wrong implementation of the $\alpha$ parameter transfers into a wrong posterior, whose inaccuracy is mainly dominated by an incorrect, too positive, skewness. The fit illustrates Eq.~(\ref{wskew}) with $\epsilon=-1$. The (c) right histogram shows the DIP test for $P^{\epsilon=0.3}(m|d)$ according to Eq.~(\ref{postb}). In this case the wrong implementation does not transfer into a significant deviation from a uniform distribution, since the mean is almost unattached.

To summarize, the DIP tests shows significant insufficiencies for $P(\sigma^2|d)$, even though it might not be the most suitable mapping to detect insufficiencies of the $\alpha$ parameter. For $P(m|d)$ the DIP test cannot detect insufficiencies. However, although one can construct infinitely many mappings onto one dimension, two natural marginalizations of the posterior onto one dimension have been sufficient to reveal the insufficiency of the implementation.

%____________________________________________________________________________________________________________________________
\bigskip
\section{Conclusion and outlook}
With the help of the introduced DIP test tools of error diagnosis it is possible to detect not only the presence of the inaccuracy but also to get an indication of its nature and the size of impact on the posterior distribution.

Furthermore, it is theoretically possible to do not only a qualitative error diagnosis, but also a quantitative study. One possibility is to consider the intersection of the distribution $P(x)$ of an insufficient posterior with the expectation value, $P(x)=1$, which encodes [in combination with the shape and slope of $P(x)$] the value of $\epsilon$. However, in reality there are combinations of different error types and numerically determined distributions are not as precise as the theoretical ones so that one might want to construct a Bayesian test for this. 

We leave the development of fully automated error detection and classification methods for future work. Inspection of the results of the DIP test by eye is already a powerful way to diagnose posterior imperfections, as we show in \cite{paper1}.

%____________________________________________________________________________________________________________________________
\begin{acknowledgments}
We want to thank Rishi Khatri and two anonymous referees for useful discussions. 
\end{acknowledgments}

%____________________________________________________________________________________________________________________________

\begin{commenta}
\appendix

\section{Calculation of special distributions of $P(x)$}
To derive the Eq.~(\ref{wvar}) and (\ref{wskew}) we start from Eq.~(\ref{proof}) and assume $P(s|d)=\mathcal{G}\left(s-\bar{s}_d,\sigma^{2}\right)$, where $\bar{s}_d$ is the data dependent maximum of the posterior. Additionally we assume $\sigma^{2}$ to be data independent for simplicity.

\begin{equation}
\begin{split}
P(x)=&\int_{-\infty}^{\infty}ds\int \mathcal{D}d~P(d)~\mathcal{G}\left(s-\bar{s}_d,\sigma^{2}\right)\\
	&\times \delta\left(x-\int_{-\infty}^s d\tilde{s}~\mathcal{G}^\epsilon\left(\tilde{s}-\bar{s}_d,\sigma^{2}\right)\right)
\end{split}
\end{equation}

\noindent Now we perform a suitable coordinate transformation, $s=s_d+\bar{s}_d$, to absorb the data dependence in the integration variable. This yields

\begin{equation}
\label{vor}
\begin{split}
P(x)=&\int_{-\infty}^{\infty}ds_d\int \mathcal{D}d~P(d)~\mathcal{G}\left(s_d,\sigma^{2}\right)~\\
	&\times \delta\left(x-\int_{-\infty}^{s_d+\bar{s}_d} d\tilde{s}~\mathcal{G}^\epsilon\left(\tilde{s}-\bar{s}_d,\sigma^{2}\right)\right)\\
	=&\int \mathcal{D}d~P(d)\int_{-\infty}^{\infty}ds_d~\mathcal{G}\left(s_d,\sigma^{2}\right)\\
	&\times \delta\left(x-\int_{-\infty}^{s_d} d\tilde{s}_d~\mathcal{G}^\epsilon\left(\tilde{s}_d,\sigma^{2}\right)\right)\\
	=&\int_{-\infty}^{\infty}ds_d~\mathcal{G}\left(s_d,\sigma^{2}\right)\delta\left(x-\int_{-\infty}^{s_d} d\tilde{s}_d~\mathcal{G}^\epsilon\left(\tilde{s}_d,\sigma^{2}\right)\right),
\end{split}
\end{equation}

\noindent where $\tilde{s}_d=\tilde{s}-\bar{s_d}$. With this expression we are able to investigate the distribution $P(x)$ for the following three cases:

\bigskip\textbf{Wrong variance.} Inserting Eq.~(\ref{wrongvar2}) into Eq.~(\ref{vor}) yields

\begin{equation}
\begin{split}
P(x)=&\int_{-\infty}^{\infty}ds_d ~\frac{1}{\sqrt{2\pi}\sigma}\exp\left(-\frac{s^2_d}{2\sigma^2}\right)\\
	&\times\delta\left(\underbrace{x-\frac{1}{2}\left[1+\text{erf}\left(\frac{s_d}{\sqrt{2}\sigma(1+\epsilon)}\right)\right]}_{=:g(s_d)}\right)\\
	=&\int_{-\infty}^{\infty}ds_d ~\frac{1}{\sqrt{2\pi}\sigma}\exp\left(-\frac{s^2_d}{2\sigma^2}\right)\times\frac{\delta\left(s_d - s_{d,0}\right)}{\left|g'(s_{d,0})\right|}\\
	=&(1+\epsilon)\exp\left(-\left[\text{erf}^{-1}\left(2x -1\right)\right]^2 \left[(1+\epsilon)^2 - 1\right]\right),
\end{split}
\end{equation}

\noindent where $s_{d,0}$ is the simple root of $g(s_d)$ and $g'(s_{d,0})=\frac{\partial g(s_d)}{\partial s_d}\bigg\vert_{s_{d,0}}$. Here we have used 

\begin{equation}
s_{d,0}=\sqrt{2}\sigma~(1+\epsilon)\text{erf}^{-1}\left(2x -1\right)
\end{equation}and
\begin{equation}
\left|g'(s_{d,0})\right|=\frac{1}{\sqrt{2\pi}\sigma(1+\epsilon)}\exp\left(-\left[\text{erf}^{-1}\left(2x -1\right)\right]^2 \right).
\end{equation}

\bigskip\textbf{Wrong skewness.} By inserting Eq.~(\ref{wrongvar3}, \ref{13}) into Eq.~(\ref{vor}) with $|\epsilon|=1$ we are yielding

\begin{equation}
\begin{split}
P(x)=&\int_{-\infty}^{\infty}ds_d ~\frac{1}{\sqrt{2\pi}\sigma}\exp\left(-\frac{s^2_d}{2\sigma^2}\right)\\
	&\times\delta\left(\underbrace{x-\frac{1}{2}\left[1+\text{erf}\left(\frac{s_d}{\sqrt{2}\sigma}\right)\right]+2T\left(\frac{s_d}{\sigma},\epsilon=\pm1\right)}_{=:h(s_d)}\right)\\
	=&\int_{-\infty}^{\infty}ds_d ~\frac{1}{\sqrt{2\pi}\sigma}\exp\left(-\frac{s^2_d}{2\sigma^2}\right)\times \sum_i \frac{\delta\left(s_d - s_{d,i}(\epsilon=\pm1)\right)}{\left|h'(s_{d,i}(\epsilon=\pm1))\right|}\\
	=&\left\{
    		\begin{array}{cc}
                		 1/2\sqrt{x} &~~~~~~~~~~\text{if}~ \epsilon=1\\
                 		 1/2\sqrt{1-x}&~~~~~~~~~~\text{if}~ \epsilon=-1
    		\end{array} 
    		\right.,
\end{split}
\end{equation}

\noindent where $s_{d,i}$ are the simple roots of $h(s_d)$, $h'(s_{d,i})=\frac{\partial h(s_d)}{\partial s_d}\bigg\vert_{s_{d,i}}$ and $T$ denotes the \textit{Owen's function} \cite{Owen}. Here we have used only two ($s_{d,1}(\epsilon=\pm1)$) of the four formal roots,

\begin{equation}
\begin{split}
s_{d,1}(\epsilon=+1)=&\sqrt{2}\sigma~\text{erf}^{-1}\left(2\sqrt{x}-1\right),\\
s_{d,2}(\epsilon=+1)=&\sqrt{2}\sigma~\text{erf}^{-1}\left(-2\sqrt{x}-1\right),\\
s_{d,1}(\epsilon=-1)=&\sqrt{2}\sigma~\text{erf}^{-1}\left(1-2\sqrt{1-x} \right),\\
s_{d,2}(\epsilon=-1)=&\sqrt{2}\sigma~\text{erf}^{-1}\left(1+2\sqrt{1-x}\right),
\end{split}
\end{equation} and
\begin{equation}
\left|h'\left(s_{d,i}(\epsilon=\pm 1)\right)\right|=\frac{1}{\sqrt{2\pi}\sigma}\exp\left(-\frac{s_{d,i}^2}{2\sigma^2}\right)\left|\text{erf}\left(\frac{s_{d,i}}{\sqrt{2}\sigma}\right)\pm1\right|.
\end{equation}
The two remaining roots, $s_{d,2}(\epsilon=\pm1)$, are not defined due to the restricted domain of $x=[0,1]$.

\bigskip\textbf{Wrong maximum position.} By inserting Eq.~(\ref{maxpos}) into Eq.~(\ref{vor}) we obtain

\begin{equation}
\begin{split}
P(x)=&\int_{-\infty}^{\infty}ds_d ~\frac{1}{\sqrt{2\pi}\sigma}\exp\left(-\frac{s^2_d}{2\sigma^2}\right)\\
	&\times\delta\left(\underbrace{x-\frac{1}{2}\left[1+\text{erf}\left(\frac{s_d-\epsilon}{\sqrt{2}\sigma}\right)\right]}_{=:q(s_d)}\right)\\
	=&\int_{-\infty}^{\infty}ds_d ~\frac{1}{\sqrt{2\pi}\sigma}\exp\left(-\frac{s^2_d}{2\sigma^2}\right)\times\frac{\delta\left(s_d - s_{d,0}\right)}{\left|q'(s_{d,0})\right|}\\
	=&\exp\left(-\frac{1}{2}\left(\frac{\epsilon}{\sigma}\right)^2-\sqrt{2}\left(\frac{\epsilon}{\sigma}\right)~\text{erf}^{-1}\left(2x -1\right)\right),
\end{split}
\end{equation}

\noindent where $s_{d,0}$ is the simple root of $q(s_d)$ and $q'(s_{d,0})=\frac{\partial q(s_d)}{\partial s_d}\bigg\vert_{s_{d,0}}$. Here we have used 

\begin{equation}
s_{d,0}=\sqrt{2}\sigma~\text{erf}^{-1}\left(2x -1\right) +\epsilon
\end{equation}and
\begin{equation}
\left|q'(s_{d,0})\right|=\frac{1}{\sqrt{2\pi}\sigma}\exp\left(-\left[\text{erf}^{-1}\left(2x -1\right)\right]^2 \right).
\end{equation}

\bigskip\textbf{Normalization.} By inserting Eq.~(\ref{norm}) into Eq.~(\ref{vor}) we obtain

\begin{equation}
\begin{split}
P(x)=&\int_{-\infty}^{\infty}ds_d ~\frac{1}{\sqrt{2\pi}\sigma}\exp\left(-\frac{s^2_d}{2\sigma^2}\right)\\
	&\times\delta\left(\underbrace{x-\frac{1}{2(1+\epsilon)}\left[1+\text{erf}\left(\frac{s_d}{\sqrt{2}\sigma}\right)\right]}_{=:p(s_d)}\right)\\
	=&\int_{-\infty}^{\infty}ds_d ~\frac{1}{\sqrt{2\pi}\sigma}\exp\left(-\frac{s^2_d}{2\sigma^2}\right)\times\frac{\delta\left(s_d - s_{d,0}\right)}{\left|p'(s_{d,0})\right|}\\
	=&1+\epsilon~~~~~\text{for}~x\in[0,1-\epsilon],
\end{split}
\end{equation}

\noindent where $s_{d,0}$ is the simple root of $p(s_d)$ and $p'(s_{d,0})=\frac{\partial p(s_d)}{\partial s_d}\bigg\vert_{s_{d,0}}$. Here we have used 

\begin{equation}
s_{d,0}=\sqrt{2}\sigma~\text{erf}^{-1}\left(2x(1+\epsilon) -1\right)
\end{equation}and
\begin{equation}
\left|p'(s_{d,0})\right|=\frac{1}{\sqrt{2\pi}\sigma(1+\epsilon)}\exp\left(-\left[\text{erf}^{-1}\left(2x(1+\epsilon) -1\right)\right]^2 \right).
\end{equation}
\end{commenta}

%____________________________________________________________________________________________________________________________
\appendix
\section{Uniformity of $P(x)$}
\textit{Proof}. We show here analytically that $P(x)=1$ if $\tilde{P}(s|d) = P(s|d)$:
\begin{equation}
\label{proof}
\begin{split}
P(x) &= \int_{-\infty}^{\infty}ds\int \mathcal{D}d ~P(x,d,s)\\
 &= \int_{-\infty}^{\infty}ds\int \mathcal{D}d ~P(x|d,s)P(d,s)\\
 &=\int_{-\infty}^{\infty}ds\int \mathcal{D}d ~P(d,s)~\delta\left(x-\int_{-\infty}^{s}ds'~P(s'|d)\right) \\
 &=\int_{-\infty}^{\infty}ds\int \mathcal{D}d~P(d)P(s|d)~\delta(x-x_{d}(s)),
\end{split}
\end{equation}
\noindent where $x_d(s):=\int_{-\infty}^{s}ds'~P(s'|d)$ and $\int \mathcal{D}d$ denotes a path integral over all possible realizations of $d$. Now we show $P(x)=1$ for $x\in[0,1]$:
\begin{equation}
\begin{split}
P&(x)=\partial_x  \int_{0}^x dx'P(x')\\
&=\partial_x \int \mathcal{D}d~P(d)\int_{-\infty}^{\infty}ds~P(s|d)\underbrace{\int_{0}^x dx'~\delta(x'-x_{d}(s))}_{\Theta(x-x_{d}(s))}\\
&=\partial_x \int \mathcal{D}d~P(d)\int_{-\infty}^{s_d(x)}ds~P(s|d) \\
&=\partial_x \int \mathcal{D}d~P(d)\underbrace{x_d(s_d(x))}_{=x} =\partial_x x\int \mathcal{D}d~P(d)\\
&=\partial_x x =1 
\end{split}
\end{equation}

\newpage
\noindent Here $s_d(x)~\text{is the inverse of}~x_d(s)$ and $\Theta$ the Heaviside step function. This inverse exists because $x_d(s)$ is strictly monotonous, unless $P(s|d)=0$ exactly for some $s$ range.~$\blacksquare$

\section{Mapping to a Gaussian}
We assume $P(x)$ to be an arbitrary one-dimensional probability distribution with related cumulative distribution, $F(X)=\int_0^X dx~ P(x)$, and $\mathcal{G}(x,1)$ to be a one-dimensional Gaussian with related cumulative distribution, $G(X)$. 

\medskip
Here, we prove that $P(y) = \mathcal{G}(y,1)$ if the coordinate transformation is given by $y(x) = G^{-1}{\left(F(x)\right)}$:

\begin{equation}
\label{trafoproof}
\begin{split}
P(y) dy = & ~P(x) dx\\
&\Leftrightarrow \\
P(y) = & ~P(x) \bigg\vert \frac{dx}{dy}\bigg\vert_{x=F^{-1}{\left(G(y)\right)}}\\
     = & ~P{\left(F^{-1}{\left(G(y)\right)}\right)} \frac{\partial F^{-1}{\left(G(y)\right)}}{\partial y}\\
     = & ~P{\left(F^{-1}{\left(G(y)\right)}\right)} \left(\frac{\partial F{\left(F^{-1}{\left(G(y)\right)}\right)}}{\partial y}\right)^{-1} \frac{\partial G(y)}{\partial y}\\
     = & ~\frac{P{\left(F^{-1}{\left(G(y)\right)}\right)}}{P{\left(F^{-1}{\left(G(y)\right)}\right)}}~ \mathcal{G}(y,1) = \mathcal{G}(y,1) \hspace{1cm}\blacksquare
\end{split}
\end{equation}

\newpage
%____________________________________________________________________________________________________________________________
\bibliography{bibliography}%apssamp}% Produces the bibliography via BibTeX.

\end{document}